\documentstyle[aps,preprint]{revtex}
\input epsf
\newcommand{\lapprox}{\stackrel{<}{\scriptstyle \sim}}
\begin{document}

\begin{titlepage}
\title{\flushright{\small LA-UR-00-600}
\vspace{-0.2in}
\flushright{\small hep-ph/0002094}
\vspace*{1in}\\
\center{\bf Relativistic Symmetry Suppresses Quark Spin-Orbit
Splitting}} \author{P.R. Page\thanks{\small \em E-mail: prp@lanl.gov},
T. Goldman\thanks{\small \em E-mail: tgoldman@lanl.gov},
and J.N. Ginocchio\thanks{\small \em E-mail: gino@t5.lanl.gov}\\{\small \em
MS B283, Theoretical Division, Los Alamos National Laboratory} \\
{\small \em Los Alamos, New Mexico 87545}}
\date{November 2000}
\maketitle

\begin{abstract}

Experimental data indicate small spin-orbit splittings in hadrons.
For heavy-light mesons we identify a relativistic symmetry that suppresses
these splittings. We suggest an experimental test in
electron-positron annihilation. Furthermore, we argue that the
dynamics necessary for this symmetry are possible in QCD.

\end{abstract}

\flushleft

\end{titlepage}

\pagebreak

\section{Introduction}

Recently, Isgur\cite{nathan} has re-emphasized the experimental fact
that spin-orbit splittings in meson and baryon systems, which might be
expected to originate from one-gluon-exchange (OGE) effects between
quarks, are absent from the observed spectrum. He conjectures that this
is due to a fairly precise, but accidental, cancellation between OGE
and Thomas precession effects, each of which has
``splittings of hundreds of MeV'' \cite{nathan}.
Taking the point of view that precise
cancellations reflect symmetries rather than accidents, we have
examined what dynamical requirements would lead to such a result.
One of us
recently observed\cite{gino} that a relativistic symmetry is the origin of
pseudospin degeneracies first observed in nuclei more than thirty years
ago\cite{kth,aa}. We find that a close relative of that dynamics can
account for the spin degeneracies observed in hadrons composed of one
light quark (antiquark) and one heavy antiquark (quark).

Below, we first elucidate the experimental evidence for small spin-orbit
splittings. Then we identify the symmetry involved in terms of potentials
in the Dirac Hamiltonian
for heavy-light quark systems, and note the relation to
the symmetry for pseudospin. We show that the former symmetry predicts
that the
Dirac momentum space wavefunctions will be identical for the
two states in the doublet, leading to a proposed experimental test.
Finally, we argue that the required
relation between the potentials may be plausible from known
features of QCD.

\section{Experimental and Lattice QCD Spectrum}

In the limit where the heavy (anti)quark is
infinitely heavy, the angular momentum of the
light degrees of freedom, $j$, is separately conserved\cite{hqet}.
The states can be
labelled by $l_{j}$, where $l$ is the orbital angular momentum of the
light degrees of freedom. In non--relativistic models of
conventional mesons the splitting between $l_{l+\frac{1}{2}}$
and $l_{l-\frac{1}{2}}$ levels, e.g.
the $p_{\frac{3}{2}}$ and $p_{\frac{1}{2}}$ or
$d_{\frac{5}{2}}$ and $d_{\frac{3}{2}}$ levels, can {\it only} arise from
spin-orbit interactions\cite{nathan}.
The $p_{\frac{1}{2}}$ level corresponds to two degenerate broad states with
different total angular momenta $J=j\pm s_Q$ (here $j=\frac{1}{2})$,
where
$s_Q$ is the spin of the heavy (anti)quark\cite{hqet}.
For example, in the case of $D$-mesons, $s_Q=\frac{1}{2}$ and
  the two states are called
  $D^{\ast}_0$ and $D_1'$. There are also two degenerate narrow
$p_{\frac{3}{2}}$ states $D_1$ and $D^{\ast}_2$\cite{hqet}.
The degenerate states
separate as one moves slightly away from the heavy quark limit, and
their spin-averaged mass remains approximately equal to the mass before
separation.

For the $D$--mesons,
the CLEO collaboration claims a broad $J^P=1^+$
state at $2461^{+41}_{-34}\pm 10 \pm 32$ MeV\cite{cleo}, belonging to
the $p_{\frac{1}{2}}$ level,
in close vicinity to the $D^{\ast}_2$ at $2459\pm 2$ MeV\cite{pdg98},
belonging to the $p_{\frac{3}{2}}$ level, indicating a remarkable
$p_{\frac{3}{2}}$-$p_{\frac{1}{2}}$ spin-orbit
degeneracy of $-2\pm 50$ MeV.
It is appropriate to extract the spin-orbit splitting this way
since: Firstly, the charm quark behaves like a heavy quark. Secondly,
the difference between the $D_1'$ and $D^{\ast}_2$ levels is the
best indicator\cite{l3} of the $p_{\frac{3}{2}}$-$p_{\frac{1}{2}}$ splitting
  in the absence of experimental data\footnote{The FOCUS collaboration
preliminarily found $D^{\ast}_0$ at a mass of $2420$ MeV \cite{dpf}.
The error on the mass was not reported.}
  on the $D^{\ast}_0$, as opposed to the
difference between the $D_1'$ and spin-averaged
$p_{\frac{3}{2}}$ level at $2446\pm 2$ MeV. Spin-averaged masses are
determined from experiment\cite{pdg98}.

For the $K$-mesons, the
$p_{\frac{1}{2}}$ level is at $1409\pm 5$ MeV, with $p_{\frac{3}{2}}$
nearby at $1371\pm 3$ MeV, corresponding to a
$p_{\frac{3}{2}}$-$p_{\frac{1}{2}}$ splitting of $-38\pm 6$
MeV. The splitting between the higher-lying $d_{\frac{5}{2}}$ and
$d_{\frac{3}{2}}$ levels is $-4\pm 14$ MeV or $41\pm 13$ MeV, depending
on how the states are paired into doublets.  These results indicate a
near spin-orbit degeneracy if the strange quark can be treated as heavy,
although it has certainly not been established that such a treatment is valid.

For $B$-mesons, both L3 \cite{l3} and OPAL \cite{opal} have performed
analyses, using input
from theoretical models and heavy quark effective theory, to
determine that the $p_{\frac{3}{2}}$-$p_{\frac{1}{2}}$
splitting is $97\pm 11$ MeV (L3) or $-109\pm 14$ MeV (OPAL).
Note that these
are {\it not} model-independent experimental results.
In the same
analyses the mass difference between the $B_2^{\ast}$ and $B_0^{\ast}$,
an approximate indicator of the $p_{\frac{3}{2}}$-$p_{\frac{1}{2}}$
splitting, is $110\pm 11$ MeV (L3) or $-89\pm 14$ MeV (OPAL).
The L3 result agrees with lattice QCD estimates
of $155^{+9}_{-13} \pm 32$ MeV\cite{gupta} and $183\pm 34$
MeV\cite{ali}. However, according to other
estimates\cite{lewis}, the splitting is less than 100 MeV, and
consistent with zero.
Recently, $31\pm 18$ MeV was calculated \cite{lewis1}.
One lattice QCD study found evidence for a change of sign in the splitting
somewhere between the charm and bottom quark masses, albeit with large
error bars \cite{hein}. A splitting of 40 MeV serves as a typical example of
model predictions \cite{godfrey}, although there is variation in the
range -155 to 72 MeV \cite{models}, summarized in ref. \cite{lewis1}.

In order to more quantitatively measure the spin-orbit splitting, define

\begin{equation}
r = \frac{(p_{\frac{3}{2}}-p_{\frac{1}{2}})}{((4 p_{\frac{3}{2}}+ 2
p_{\frac{1}{2}})/6 - s_{\frac{1}{2}})},
\end{equation}
where all entries refer to masses.
The experimental data on $D$, $K$ and $B$ mesons
give respectively $r=0.00\pm 0.10, \; -0.06\pm 0.00$ and $0.23 \pm 
0.04$ (L3) or $-0.23\pm 0.03$ (OPAL).
For the Dirac equation with arbitrary vector and scalar Coulomb potentials,
the only cases for which the relevant analytic solutions are known,
$-0.7 \lapprox r \lapprox 0.6$. It is hence evident that the spin-orbit
splittings extracted from experimental results are indeed small.

There is also evidence in light quark mesons and baryonic systems that the
spin-orbit interaction is small\cite{nathan}. In
non-relativistic models, meson and ``two-body'' baryon spin-orbit
interactions are related and, for a specific class of
baryons, the spin-orbit interaction is small for exactly the same
reasons that it is small in mesons\cite{nathan}.

\section{A Dynamical Symmetry for the Dirac Hamiltonian}

If we consider a system of a (sufficiently) heavy antiquark (quark) and
light quark (antiquark), the dynamics may well be represented by the
motion of the light quark (antiquark) in a fixed potential provided by
the heavy antiquark (quark). Let us assume that both vector and scalar
potentials are present. Then the Dirac Hamiltonian describing the motion
of the light quark is
\begin{equation}
H = {\vec{\alpha}} \cdot {\vec p} + \beta (m + V_S) + V_V + M , \label{dirac}
\end{equation}
where we have set $\hbar = c =1$, ${\vec \alpha}$, $\beta $ are the
usual Dirac matrices, $\vec p$ is the three-momentum, $m$ is the mass
of the light quark and $M$ is the mass of the heavy quark.

This one quark Dirac Hamiltonian follows from the two-body Bethe-Salpeter
equation in the equal time approximation, the spectator (Gross) 
equation with a simple
kernel, and a two
quark Dirac equation, in the limit that $M$ is
large\cite{bethe,gross,pik}. If the vector potential, $V_{V}(\vec r)$,
is equal to the scalar potential plus a constant potential, $U$, which
is independent of the spatial location of the light quark relative to
the heavy one, i.e., $V_{V}(\vec r) = V_{S}(\vec r) + U$, then the Dirac
Hamiltonian is invariant under a spin symmetry\cite{bell,ami},
$[\,H\,,\, {\hat { S}}_i\,] = 0$, where the generators of that symmetry
are given by,
\begin{equation}
{\hat {S}}_i = \left ( {{\hat {s}_i} \atop 0 } { 0 \atop {
{\hat{\tilde s}}_i}}\right ).
\label{sgen}
\end{equation}
where ${\hat s}_i = \sigma_i/2$ are the usual spin generators,
$\sigma_i$ the Pauli matrices, and ${\hat {\tilde s}}_i = U_p\ {\hat
{s}}_i \ U_p $ with $U_p = \, {{ \vec \sigma\cdot \vec p} \over p}$.
Thus Dirac eigenstates can be labeled by the orientation of the spin,
even though the system may be highly relativistic, and the eigenstates
with different spin orientation will be degenerate.

For spherically symmetric potentials, $V_{V}(\vec r) = V_{V}(r),\;
V_{S}(\vec r) = V_{s}( r)$, the Dirac Hamiltonian has an additional
invariant algebra; namely, the orbital angular momentum,
\begin{equation}
{\hat {L}}_i = \left ( {{\hat {\ell}_i} \atop 0 } { 0 \atop {
{\hat{\tilde \ell}}_i}}\right ),
\label{jgen}
\end{equation}
where ${\hat {\tilde \ell}}_i = U_p\, {\hat \ell}_i$ $U_p$ and
${\hat\ell}_i = (\vec r \times \vec p)_i$. This means that the Dirac
eigenstates can be labeled with orbital angular momentum as well as spin, and
the states with the same orbital angular momentum projection will be
degenerate. Thus, for example, the $n_r\ p_{1/2}$ and $n_r\ p_{3/2}$
states will be degenerate, where $n_r$ is the radial quantum number.

Thus, we have identified a symmetry in the heavy-light quark system
which produces spin-orbit degeneracies independent of the details of
the potential. If this potential is strong, the heavy-light quark
system will be very relativistic; that is, the lower component for the
light quark will be comparable in magnitude to the upper component of
the light quark.
It is remarkable that non-relativistic behaviour of energy levels can
arise for such fully relativistic systems.

This symmetry is similar to the relativistic symmetry\cite{gino}
identified as being responsible for pseudospin degeneracies observed in
nuclei\cite{kth,aa}. In contrast to spin symmetry, pseudospin symmetry
has the pairs of states $((n_r-1)s_{1/2}, n_rd_{3/2})$,\\
$((n_r-1)p_{3/2}, n_rf_{5/2})$, etc. degenerate, making the origin of
this symmetry less transparent.  The pseudospin generators are
\begin{equation}
{\hat {\tilde S}}_i = \left ( {{\hat {\tilde s}_i} \atop 0 } { 0
\atop { {\hat { s}}_i}}\right ).
\label{gen}
\end{equation}
For pseudospin symmetry, the nuclear mean scalar and vector potential
must be equal in magnitude and opposite in sign, up to a constant,
$V_{V} = -V_{S}+U$.  Relativistic mean field representations of the
nuclear potential do have this property; that is,
$V_{S}~\approx~-V_{V}$\cite{dirk,mad}.  We will return later to the
question of whether the relation $V_{V} = V_{S} + U$ arises in QCD.

It has previously been observed that pseudospin symmetry improves with
increasing energy of the states, for various potentials\cite{gino}. A
similar behaviour may be expected for spin symmetry, consistent with the
experimental observations that spin--orbit splittings decrease for
higher mass states\cite{nathan,pdg98}.

The Dirac Hamiltonian (\ref{dirac}) encompasses the effects
of the OGE and Thomas
precession spin-dependent terms customarily included in non-relativistic
models\cite{nathan}.

\section{Experimental Test}

In the spin symmetry limit, the radial wavefunctions of the upper
components of the Dirac wavefunction of the two states in the spin
doublet will be identical, behaving ``non-relativistically'',
whereas the lower components will have
different radial wavefunctions.  This follows from the form of the spin
generators given in Equation (\ref {sgen}). The $(1,1)$ entry of the
operator matrix is simply the non-relativistic spin operator which
relates the upper component of the Dirac wavefunction of one state in
the doublet to the upper component of the other state in the doublet.
Since this operator does not affect the radial wavefunction, the two
radial wavefunctions must be the same. By contrast, the lower component
wavefunction is operated on by $U_p$ which does operate on the radial
wavefunction because of the momentum operator.

As an example, we show in Figure 1 the upper and lower components for
Dirac wavefunctions of the $p_{1/2} - p_{3/2}$ doublet. The
scalar and vector potentials were determined by matching the available
spectral data of
the $D$-mesons, assuming a $p_{\frac{3}{2}}-p_{\frac{1}{2}}$ splitting at
the lower end of the range defined by the experimental value of
$-2\pm 50 $ MeV. This maximizes the wavefunction differences. In this 
realistic case, $V_V \approx V_S + U$, so the
radial wavefunctions for the upper components are not exactly identical
but are very close, whereas the radial wavefunctions for the lower
components are very different.

Likewise the momentum space wavefunctions for the upper
components will be very similar, as seen in Figure 2, again because
the spin operator does not affect the wavefunction.  However,
since $U_p$ depends only on the angular part of the momentum, $ \hat p
= {\vec p \over p}$, it does not affect the radial
momentum space wavefunction. In Figure 2 we see that the radial
momentum space wavefunctions are very similar for the lower components as well.
This prediction of the symmetry can be tested in the following experiment.

The  annihilation $e^+e^- \rightarrow
D^{\ast}_0 D^{\ast}_0$, $D^{\ast}_0 D^{\ast}_2$
and $D^{\ast}_2 D^{\ast}_2$ allows for the extraction of the
$D^{\ast}_0$ and $D^{\ast}_2$ electromagnetic static form factors and the
$D^{\ast}_0$ to $D^{\ast}_2$ electromagnetic transition form factor.
The photon interaction ensures that all
radial wavefunctions of the light quark are accessed.
When spin symmetry is realised,
there are only two independent radial momentum space
wavefunctions, which should enable
the prediction of one of the three form factors in terms of the other two.
This should enable the verification of the predictions of spin symmetry.
On the other hand, non-relativistic models, with no lower components for
the wavefunctions, have only one independent radial wavefunction, which will
lead to the prediction of two of the form factors in terms of
the remaining
one. This might be
too restrictive. The proposed experiment can be carried
out at the Beijing Electron Positron Collider at an energy of
approximately 1 GeV above the $\psi(4040)$
peak in the final state $DD\pi\pi$.

An equivalent experiment for K-mesons would involve detection of the
$KK\pi\pi$ final state, which has already been measured\cite{dm2}.
The wavefunctions of $K$-mesons fitting the experimental spectrum
show similar behaviour to the $D$-mesons, with
the $p_{\frac{3}{2}}$ and $p_{\frac{1}{2}}$ wavefunctions even more similar
than in Figures 1a and 2.

If B-mesons do also exhibit spin symmetry, one can do equivalent
experiments around 1 GeV above the $\Upsilon$(3S) peak at the SLAC, KEK or CESR
B-factories.

\section{QCD Origins}

If such a dynamical symmetry can explain the suppression of spin-orbit
splitting in the hadron spectrum, the question remains as to why it
might be expected to appear in QCD. To address this, we first recall
the ongoing  argument as to whether confinement corresponds to a vector
or scalar potential\cite{refs}. The first natural expectation was that
confinement reflected the infrared growth of the QCD coupling constant,
enhancing the color-Coulomb interaction at large distances, see
e.g. Ref.(\cite{BuchTye}).
An involved two- (or multi-) gluon effect has been proposed\cite{goldh}
to account for the origin of  a scalar confining potential.

The existence of one or the other of these vector and
scalar potentials is not necessarily exclusionary -- they may both be
realised.  The arguments in Ref.(\cite{Goldman}) suggest further that
they are related, with the scalar exceeding the vector by an amount
which may be approximately constant as one saturates into the linear
confining region at large separations. We very briefly reiterate the
basic argument of Ref.(\cite{Goldman}) here.

The starting point is to accept the standard approach\cite{BuchTye}
that renormalization-group-improved single-gluon-exchange produces a
linearly increasing vector potential between a quark and an antiquark.
One then considers what to expect for multiple gluon exchange, starting
with two gluons. Since two gluons are attracted to each other in a
color singlet channel, and also have a zero mass threshold (as for
massless quark-antiquark pairs), it is reasonable to conclude that a
(Lorentz and color) scalar gluonic condensate develops, along with a
mass gap for a glueball state. These developments are indeed
observed in lattice QCD calculations.

Ref.(\cite{Goldman}) goes on to argue that
renormalization-group-improved single-glueball-exchange involves the
square of the QCD coupling and so, despite the massiveness of the
object exchanged, also leads to a (now scalar) confining potential
between quarks and antiquarks.
This further implies that the ratio of the slopes
of the two potentials in their common linear (confining) region is
given by the square of the ratio of the QCD scale for growth of the
coupling constant to the value of the mass gap of the condensate
formation.  This ratio may be expected to be of order one as both
quantities are determined by the underlying QCD scale.

If the two potentials do indeed have similar slopes in the region
outside that dominated by the color Coulomb interaction,
they would necessarily differ only by an approximately constant value, in
that region.  Thus, the origin of the dynamical symmetry may not be
unreasonable, and may indeed be a natural outcome of non-perturbative QCD.

On the other hand, {\em identically} equal vector and scalar
potentials, except for a constant difference, would appear to be coincidental.
An ameliorating effect is that to produce an approximation to the spin
symmetry of Eq. (2) this condition
need only hold in regions where the wavefunctions are substantial.

The determination of QCD potentials, from models like the minimal area
law, stochastic vacuum model, or dual QCD, and from lattice QCD, is
hampered by the problem of rigorously defining the concept of a
potential from QCD when one quark is light. It suffices to say that there
is no agreement on the mixed Lorentz character of the potential even
between two heavy quarks\cite{ebert}, where the potential can be
rigorously defined, although lattice QCD results are
consistent with simply a vector Coulomb and scalar linear
potential\cite{bali}.

\section{Summary}

The observation of ``accidental'' spin-orbit degeneracies observed in
heavy-light quark mesons can be explained by a relativistic symmetry of
the Dirac Hamiltonian which occurs when the vector and scalar
potentials exerted on the light quark by the heavy antiquark differ
approximately by a constant, $V_V \approx V_S + U$.
Conversely, if future experiments determine that spin-orbit
splittings are small not only for the lowest excited states in mesons
but are small throughout the meson spectrum, this experimental fact
dictates that the effective QCD vector and scalar potentials between
a quark and antiquark
  are approximately equal up to a constant, which would be a
significant observation about the nature of non-perturbative QCD.
Furthermore, the approximate
symmetry predicts that the spatial Dirac wavefunction for the spin
doublets will be approximately equal in momentum space, a feature which
can be tested in electron-positron annihilation.
We have argued
that $V_V \approx V_S + U$ may occur in QCD, particularly for regions
of space dominated by the light quark wavefunction.
Work is in progress to extend this symmetry to purely light quark systems.

This research is supported by the Department of Energy under contract
W-7405-ENG-36.

\newpage

\begin{center}
{\bf FIGURES}
\end{center}

Figure 1: (a) The square of the Dirac radial wavefunction of the upper
component times $r^2$. (b) The square of the Dirac radial wavefunction of
the lower component times $r^2$. $p_{\frac{3}{2}}$ is the solid line and
$p_{\frac{1}{2}}$ is the dashed line. Note that the lower component 
is comparable
to the upper component. The wavefunctions are solutions of the Dirac
equation (see Eq. (\ref{dirac})) with Coulomb potentials $V_S(r) = 
\frac{\alpha_S}{r}+U_S$ and
$V_V(r) = \frac{\alpha_V}{r}+U_V$, where $\alpha_S = -1.279,\; U_S = 506$
MeV, $\alpha_V = -0.779,\; U_V = 515$ MeV, $m=330$ MeV and
$M = 1480$ MeV. This corresponds to a $p_{\frac{3}{2}}-p_{\frac{1}{2}}$
splitting of -52 MeV.

Figure 2: (a) The square of the Dirac momentum space wavefunction of the
upper component times $q^2$. (b) The square of the Dirac momentum space
wavefunction of the lower component times $q^2$. Other conventions are the
same as in Figure 1.

\epsfbox{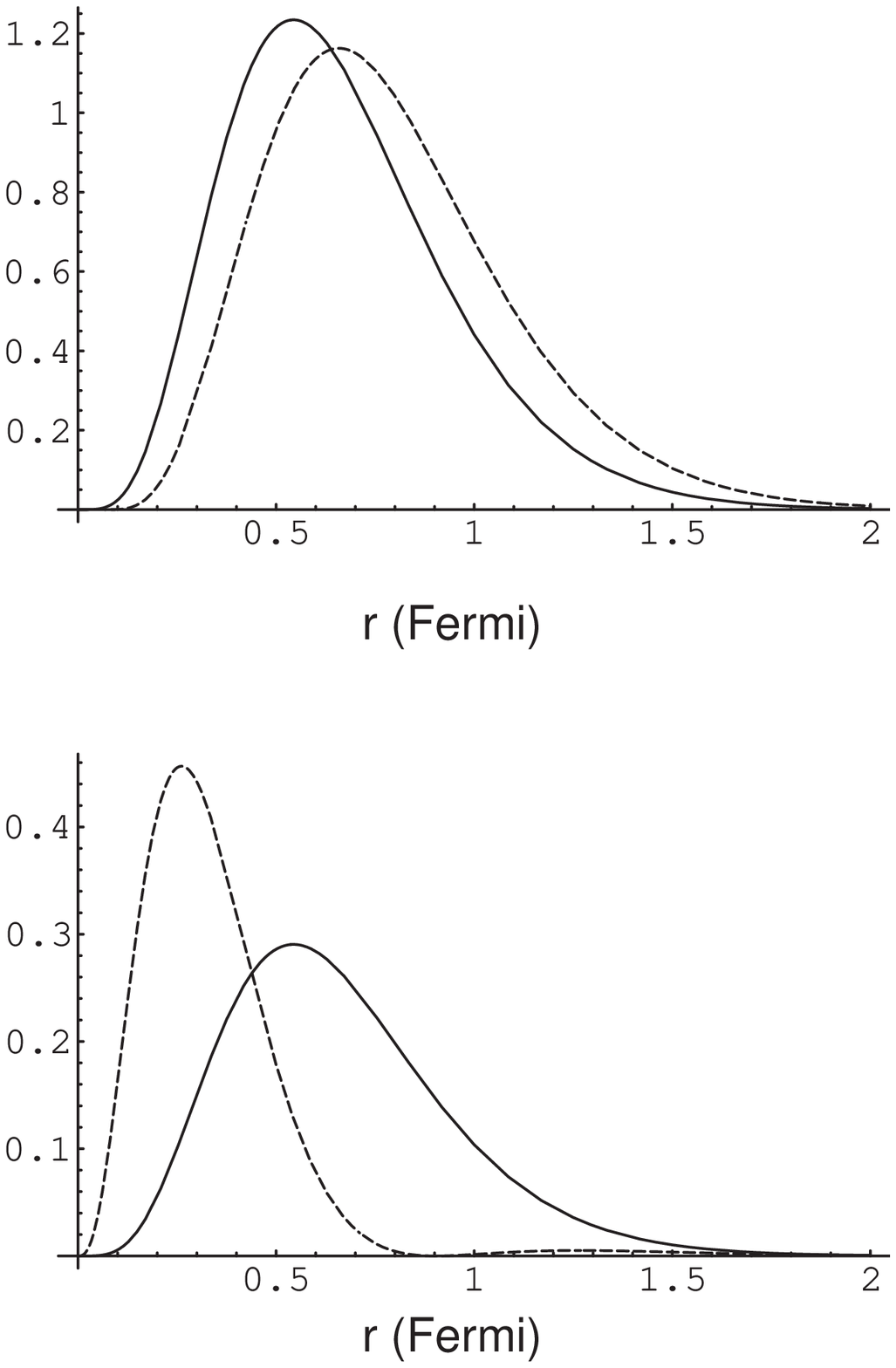}

\newpage

\epsfbox{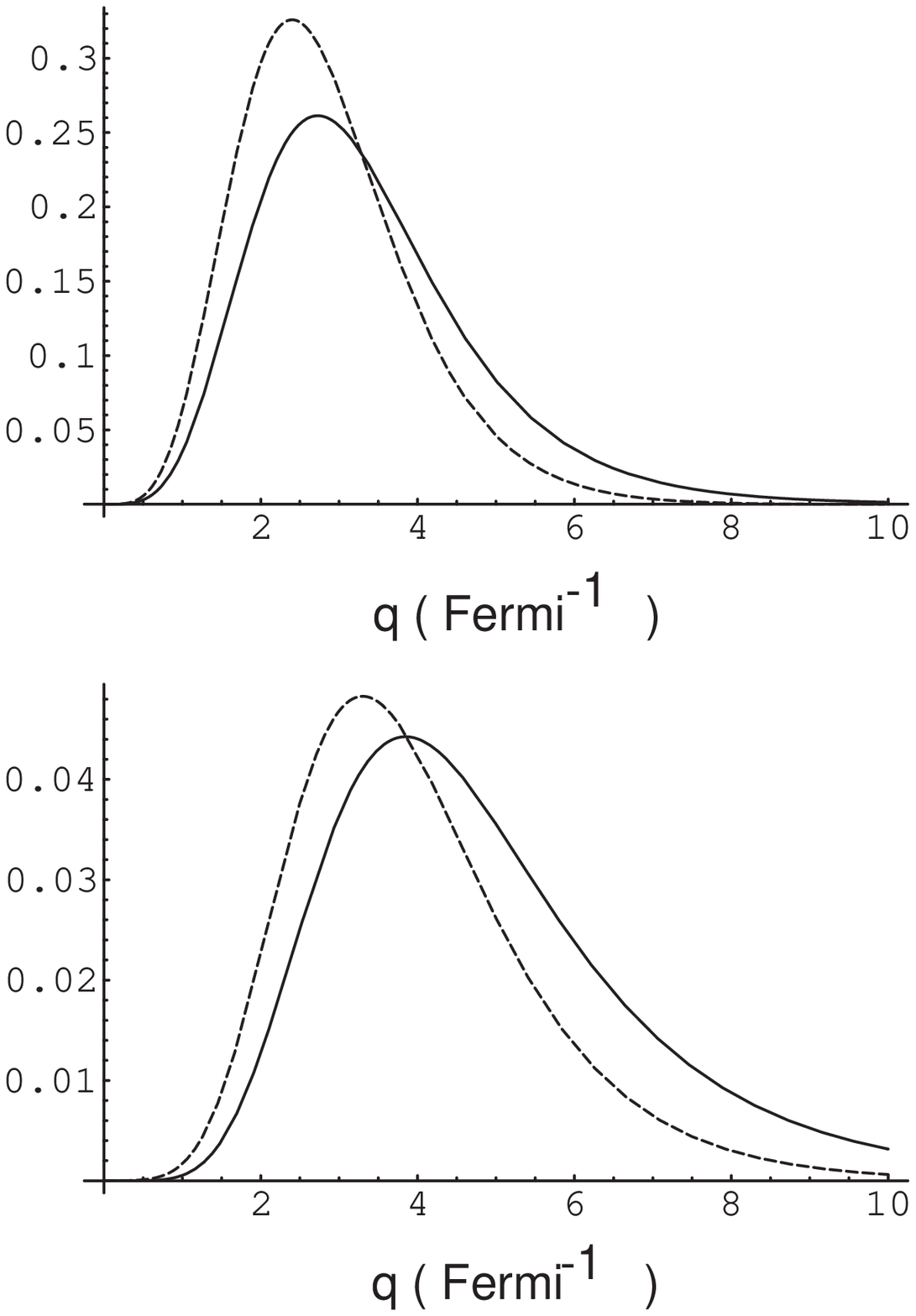}



\end{document}